\documentclass[twocolumn,aip,rsi,reprint,bibnotes]{revtex4-1}

\usepackage[T1]{fontenc}

\usepackage{natbib}
\usepackage{graphicx}
\usepackage{bm}
\usepackage{color}
\usepackage{amsmath}

\newcommand\unit[2]{\ensuremath{#1~\mathrm{{#2}}}}

\begin{document}

\title{Implementation of a stable, high-power optical lattice for quantum gas microscopy}

\author{A. Mazurenko}
\author{S. Blatt}
\altaffiliation{
  Present address: Max-Planck-Institut f{\"u}r Quantenoptik,
  Hans-Kopfermann-Stra{\ss}e 1, 85748 Garching, Germany}
\author{F. Huber}
\altaffiliation{
  Present address: IPG Photonics Corporation, Oxford, MA 01540}
\author{M. F. Parsons}
\author{C. S. Chiu}
\author{G. Ji}
\author{D. Greif}
\author{M. Greiner}
\affiliation{
  Department of Physics, Harvard University,
  Cambridge, Massachusetts, 02138, USA}

\date{\today}
\begin{abstract}
We describe the design and implementation of a stable high-power \unit{1064}{nm} laser system to generate optical lattices for experiments with ultracold quantum gases. The system is based on a low-noise laser amplified by an array of four heavily modified, high-power fiber amplifiers. The beam intensity is stabilized and controlled with a nonlinear feedback loop.
Using real-time monitoring of the resulting optical lattice, we find the stability of the lattice site positions to be well below the lattice spacing over the course of hours.
The position of the harmonic trap produced by the Gaussian envelope of the lattice beams is stable to about one lattice spacing and the long-term (six-month) relative RMS stability of the lattice spacing itself is 0.5\%.
\end{abstract}
\maketitle

Optical lattices holding arrays of ultracold atoms have become a powerful experimental platform for quantum simulations~\cite{Friedenauer2008, Kim2010, Struck2011, Simon2011, Yan2013, Greif2013, Murmann2015, Drewes2017}.
The recent advent of quantum gas microscopy~\cite{Bakr2009, Sherson2010, Haller2015, Cheuk2015, Parsons2015, Edge2015, Omran2015, Greif2016, Cheuk2016, Parsons2016, Boll2016, Cheuk2016a, Brown2017} has enabled studies with unprecedented control over individual atoms in many-body systems governed by the Hubbard Hamiltonian~\cite{Hubbard1963}, resulting in extremely low energy scales~\cite{Mazurenko2017, Chiu2018}.
However, this progress has placed ever more stringent technical constraints on the lasers used to trap and manipulate the atomic systems.
Even if the systems thermalize, heating rates can limit the interrogation time available to probe low-temperature states~\cite{Blatt2015}.
In particular, experiments that rely on a small number of atoms held by multiple traps~\cite{Preiss2015, Choi2016, Mazurenko2017, Chiu2018} place stringent requirements on the relative positional stability of the traps, to ensure that the operations performed by the traps have adequate fidelity.

In this work we discuss major improvements to our high power optical lattice system~\cite{Blatt2015} and show levels of intensity noise below \unit{-115}{dBc/Hz} between \unit{1}{kHz} and \unit{3}{MHz} Fourier frequency.
In addition to the intensity stability, we demonstrate a short-term (shot-to-shot) stability of the lattice position at 1.4\% of the optical lattice spacing, and a long-term (six month) relative stability of 0.5\% RMS of the lattice spacing itself.

For context, typical high-power, low-noise \unit{1064}{nm} laser systems are based on amplification of a seed laser with either a fiber or a solid-state amplifier.
For example, the laser system used for gravitational wave detection at advanced LIGO exceeds \unit{-150}{dBc/Hz} in the \unit{10}{Hz} -- \unit{1}{kHz} frequency range for a constant output power~\cite{kwee12}.
In this work, we achieve approximately \unit{-120}{dBc/Hz} intensity stability in the \unit{1}{kHz} -- \unit{2}{MHz} frequency range, and maintain this performance over six orders of magnitude in output power. Our modifications to the fiber amplifier hardware allow us to rival and exceed the noise performance of state-of-the-art commercial fiber amplifiers~\cite{guiraud16}.

\begin{figure}[t]
  \begin{center}
    \includegraphics[width=0.8\columnwidth]{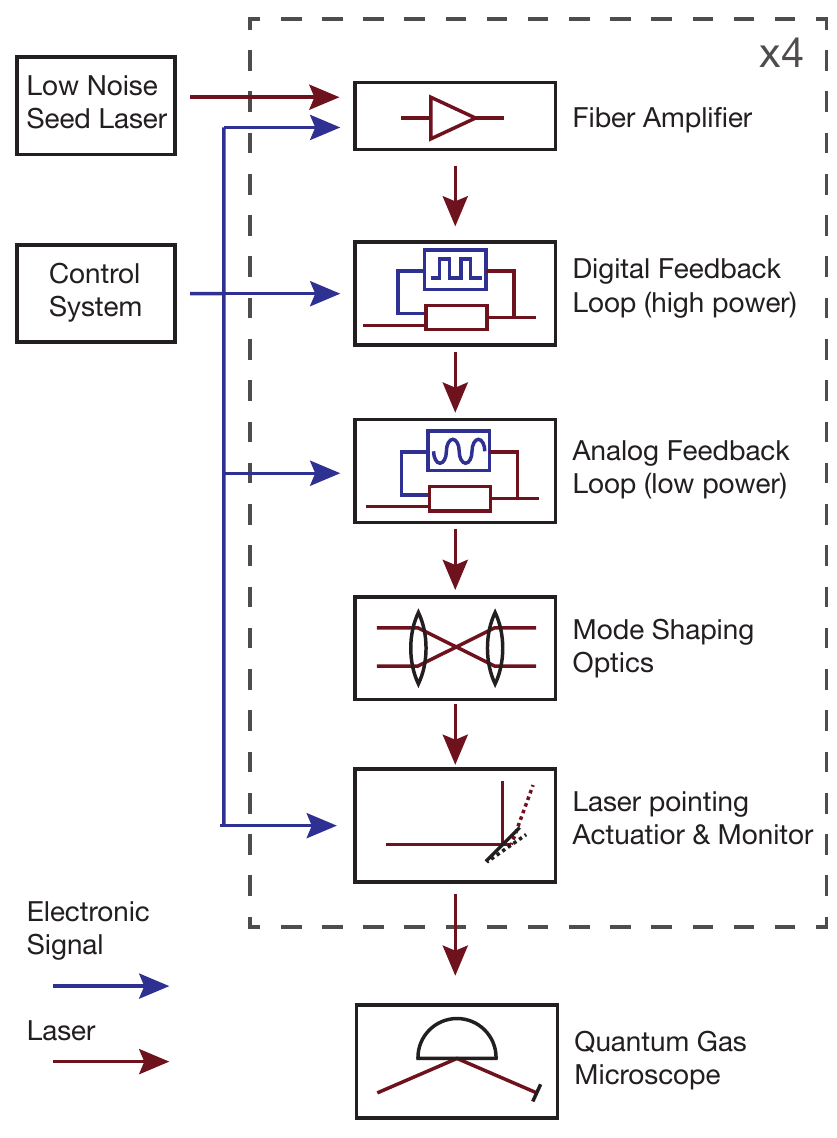}
    \caption{\textbf{High level overview} showing the major components of the lattice laser feedback system.}\label{fig:high_level}
  \end{center}
\end{figure}

The architecture of the system, shown in Fig.~\ref{fig:high_level}, is based on a single low-power, low-noise seed laser that supplies light to an array of four high-power fiber amplifiers (one for a transport beam~\cite{Huber2014}, and one each for the $X,Y,Z$ lattices~\cite{Blatt2015,Parsons2016}).
These amplifiers generate sufficient light to perform site-resolved imaging~\cite{Parsons2015}, but also substantially increase the level of intensity and pointing noise.
The fiber amplifier output is controlled by two feedback loops, which stabilize the light in low power and high power regimes, respectively, as discussed later.
After intensity stabilization, the laser beam is mode-shaped to a desired waist.
The optical system is engineered to be passively stable in terms of pointing, but the position and phase of the optical lattice are monitored in real time and beam pointing can be controlled with remotely actuated mirrors.

In the following we give an in-depth technical description of the optical lattice setup we developed. We start with the laser system itself which includes the seed laser, amplification and optical setup for controlling the laser beams. In the next section we provide a detailed description of the two-stage laser power feedback system. In the last two sections we conduct a performance analysis of the short- and long-term stability of the optical lattices produced by this system.
Naturally, the laser system presented here is tailored to suit the experiment's specific lattice geometry~\cite{Huber2014} and exact requirements.
We expect, however, that both the specific engineering choices and the general design practices detailed in this work will be widely applicable to experimentalists working with low-noise and stable optical lattices used for quantum many-body physics, precision measurement, and quantum engineering.

\section{Laser system}
\subsection{Seed laser system}
The lattice laser system is seeded by an Innolight Mephisto laser providing approximately \unit{2}{W} of \unit{1064}{nm} light.
This commercial laser uses a non-planar ring oscillator (NPRO) design that ensures superb phase and intensity stability.
This source is split and coupled into four single-mode, polarization-maintaining fibers (Thorlabs PM980-XP), each connected to the input of a fiber amplifier.
Three of these fiber amplifiers are used to produce the 3D optical lattice~\cite{Blatt2015,Parsons2016}, and a fourth is used for transport~\cite{Huber2014}.
To avoid interference of the lattice beams, two of the beam paths pass through acousto-optic modulators (AOMs), which detune the lattice frequencies from each other.

\begin{figure*}
  \begin{center}
    \includegraphics{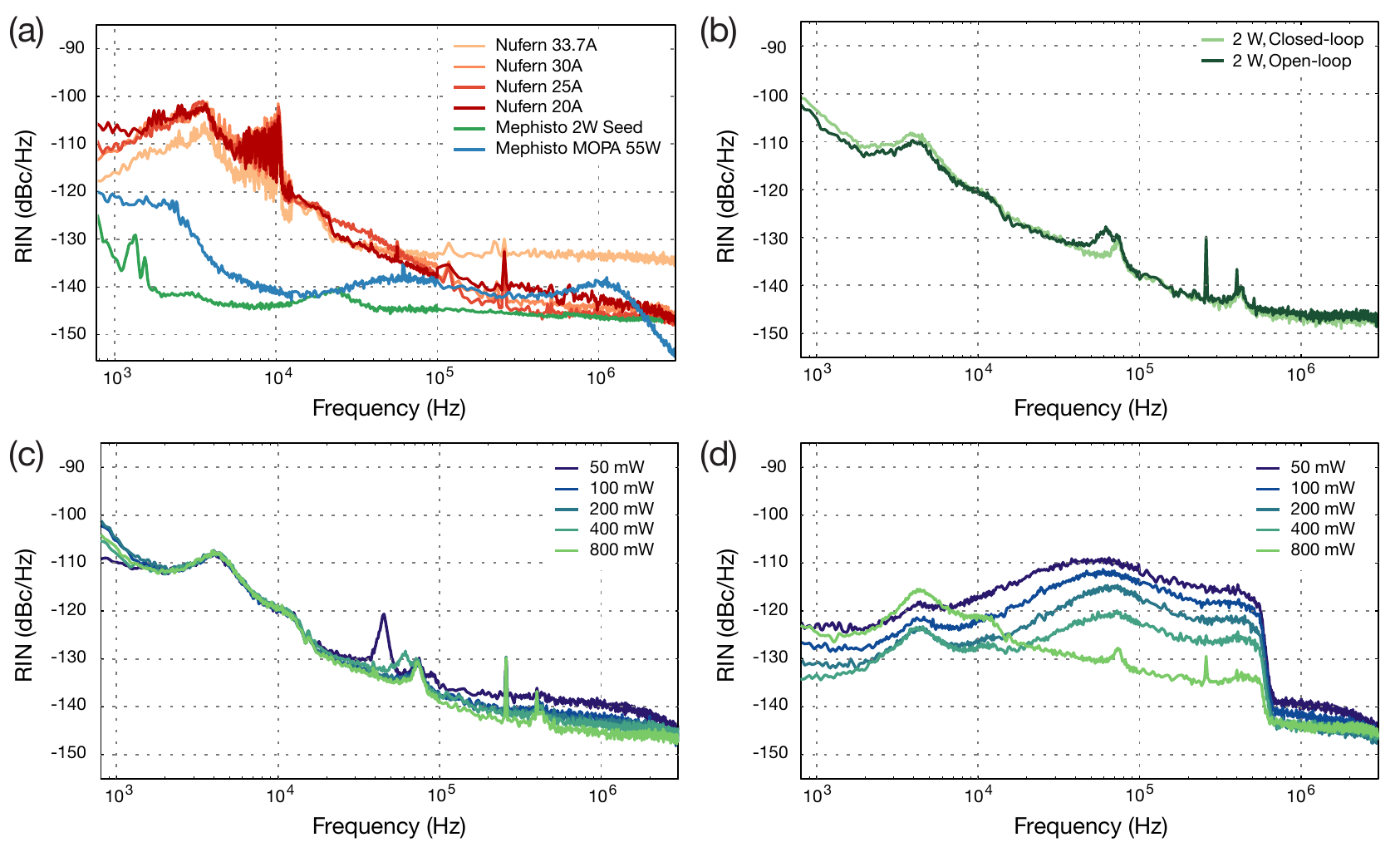}
    \caption{\textbf{Laser noise of the lattice system.} (a) Shows the laser noise of unmodified and free-running Nufern fiber amplifiers and the Mephisto seed, as well as an independent Mephisto MOPA 55W (not used for the optical lattice presented in this manuscript). (b) Shows the open-loop and closed-loop behavior of the high-power feedback loop using the  modified fiber amplifiers, as described in the main text. (c) Shows the open-loop behavior of the low-power feedback loop using the modified fiber amplifiers at various constant powers. (d) Closed-loop behavior of the low-power feedback loop (without additional low-pass filtering), measured on the same setup as panel (c).}\label{fig:noises}
  \end{center}
\end{figure*}

\subsection{Fiber amplifier system}
Quantum gas microscopy of $^6$Li requires single-site trap frequencies of approximately \unit{1}{MHz} to be in the Lamb-Dicke regime necessary for Raman sideband cooling~\cite{Parsons2015}.

Such trap frequencies may only be attained using tightly focused, high-power beams.
In practical terms, this requires approximately \unit{20}{W} per lattice axis for a beam waist of around \unit{80}{\mu m} and the four-fold enhancement in lattice depth provided by the reflection from the superpolished substrate~\cite{Huber2014}. We achieve such high laser powers using Yb-doped fiber amplifiers (Nufern NuAmp SUB1174-34).
These amplifiers can amplify \unit{150}{mW} of seed light to up to approximately \unit{45}{W}. They use a length of doped fiber pumped with \unit{808}{nm} light from fiber coupled diode sources to amplify the \unit{1064}{nm} seed light.
The amplification is performed using a two-stage design, where the stages are controlled and powered using separate power supplies and electronics.

The NuAmp suffers from several technical problems.
It is controlled via a USB interface, which complicates communications with an array of these devices.
Worse yet, the digital control system causes ground loops that add noise and interference to the analog systems.
This added noise can be seen in Fig.~\ref{fig:noises}, where the fiber-amplified light exhibits many more spurs in the intensity noise spectrum than the seed laser, particularly near \unit{10}{kHz}.
We solve this problem by replacing the USB control board with a custom-designed solution that is compatible with our laboratory control system and engineered to avoid ground loops.

Further sources of technical noise arise from the design of the amplifier. There are two power supplies in the fiber amplifier: (1) a general purpose power supply for the control electronics and the first, low-power amplification stage and (2) a high-power supply for the second, high-power stage.
Despite the presence of a water-cooled cold-plate for the gain fiber and pump diodes, the high power supply is air-cooled by an internal fan with variable frequency.
This fan adds noise at acoustic frequencies and degrades pointing stability by increasing vibration of the output fiber tip.
To reduce acoustic coupling and reduce switching power supply noise, both power supplies were removed from the fiber amplifier chassis, and the low-power supply was replaced by a linear power supply (Acopian A24MT350M).
Further, a high current low-pass filter (MPE DS26387) was added to the high-power supply (Lumina LDD-600-60-10-5VP/M).

These upgrades serve to suppress many of the noise spurs measured in the fiber amplifier RIN.
The effects can be seen when comparing Fig.~\ref{fig:noises}(a) and (b), particularly in the region of approximately \unit{10}{kHz} Fourier frequency.
The measurements in panels (a)-(d) were carried out on a single fiber amplifier, prior to and following the modifications. The initial and final noise spectra were verified to be consistent between amplifiers coming from the same manufacturing batch.
The figure is based on fiber amplifiers from a relatively new batch, which even in its unmodified configuration does not exhibit a large switching supply spur.
This is not the case for certain older batches, where the spur is approximately \unit{30}{dB} above the background.

The high optical powers supplied by the fiber amplifiers raise two engineering difficulties:
\begin{enumerate}
  \item The fiber amplification process itself is susceptible to many technical noise sources, such as acoustic coupling of the fiber to the environment and electrical design of the amplifier;
  \item Stimulated Brillouin scattering (SBS) places a hard lower bound on the relative intensity noise (RIN) floor~\cite{Agrawal2013}.
\end{enumerate}

The SBS noise is fundamental to the fiber amplifier architecture and cannot be entirely avoided. However, since it is a function of fiber length and optical power~\cite{Agrawal2013}, it can be minimized. To this end, we use custom-ordered amplifiers with a \unit{50}{cm} long output fiber, compared to the standard length of a few meters.
In addition, we use the lowest pump diode current that yields enough power to perform imaging.
The rise of the noise floor as a function of the pump diode current can be seen in Fig.~\ref{fig:noises}(a) at Fourier frequencies above \unit{100}{kHz}.

The operation of the Nufern fiber amplifiers can be further optimized by finely controlling the cooling water temperature.
The water temperature tunes the wavelength of the light supplied by the diodes that optically pump the gain fiber.
Optimizing the temperature is advantageous because (1) the maximum output power of the fiber amplifier is increased due to the improved efficiency of pump light conversion and (2) the lifetime of the amplifier is increased because less stray pump light needs to be absorbed when the pump light is separated from the output, reducing the thermal load.

This is possible because the pump diodes are fixed directly to the water-cooling plate without additional temperature regulation.
Adding a thermoelectric cooler and regulating to the optimal wavelength is possible, but technically challenging (requiring cooling power of $\approx$\unit{100}{W} for each of the three pump diodes).
We use a single, dedicated heat exchanger (TermoTek P21518970) to supply cooling water to an array of four fiber amplifiers.
We tune the water temperature such that the average power supplied by the amplifiers supplying the $X$ and $Y$ lattices is optimized, while simultaneously keeping all reported amplifier temperatures below \unit{30}{^\circ{}C}, as recommended by a Nufern support engineer.
Experimentally, we have found that more power can be gained by running the amplifiers at even hotter temperatures, possibly at the expense of a shortened lifetime.

\subsection{Optical system}
The free-space propagation of the fiber amplifier output presents several technical challenges and constraints:
\begin{enumerate}
  \item The high optical power poses a significant danger to users and equipment.
  \item The high optical power is sufficient to induce substantial thermal lensing in some optics.
  \item The positional fluctuations of the waist at the atoms cannot be allowed to exceed a few microns between experimental realizations.
  \item By construction, the beam is retro-reflected, threatening the lifetime of the amplifier.
  \end{enumerate}

\begin{figure*}[t]
  \begin{center}
    \includegraphics{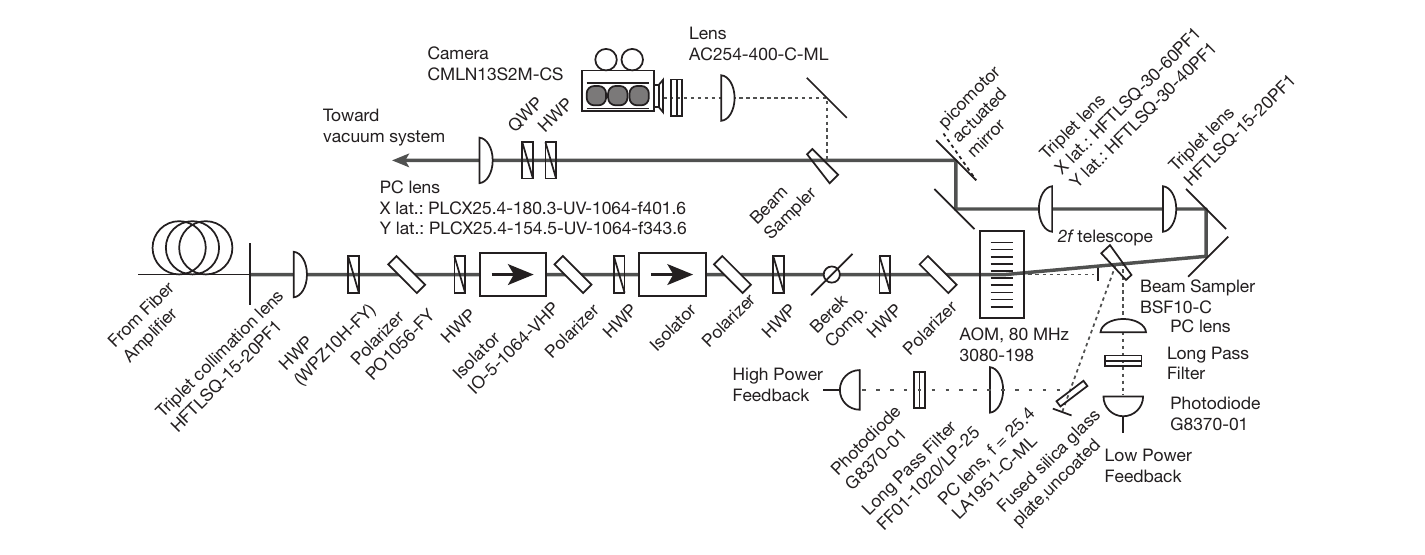}
    \caption{\textbf{Optical layout} of the optical lattice system for beam shaping and modulation.}\label{fig:optical_layout}
  \end{center}
\end{figure*}

To address these challenges, we use a carefully engineered optical system shown in Fig.~\ref{fig:optical_layout}.
This system collimates and polarizes the beam, controls its intensity and ensures good pointing stability.
Due to the high laser power, fused silica glass and ion-beam-sputtered (IBS) coatings are used wherever possible.
Since reflections pose a danger to the users and equipment, all undesired beams above \unit{100}{mW} are directed to water cooled beam-dumps (to prevent temperature gradients on the table) placed as far as practical from critical beam paths.
Weaker beams are dumped on uncooled diffusive beam catchers.

The large-diameter fiber tip at the output of the fiber amplifier is mounted in a monolithic mount made from copper.
The beam is collimated using an $f=\unit{20}{mm}$ fused-silica, air-spaced triplet collimator (Opto-Sigma HFTLSQ-15-20PF1) angled with respect to and centered on the angle-cleaved fiber tip.
The polarization is set using an IBS-coated thin-film plate polarizer (Precision photonics PO1056-FY), with the added benefit of rejecting the undesired cladding modes of the fiber, which carry about \unit{1}{W} of light. A thin-film polarizer was used to minimize the number of normal and backreflecting surfaces near the output of the fiber in addition to providing excellent polarization performance.
The fact that the lattice is retro-reflected, combined with the observation that optical isolation falls with applied optical power~\cite{Yoshida1999} means that two stages of optical isolators must be used.
The optical isolators are based on \unit{5}{mm} aperture Tb$_3$Ga$_5$O$_{12}$ (TGG) isolators (Thorlabs IO-5-1064-VHP), where we replaced the default polarizers (BK7 Brewster plates) by IBS-coated fused silica plate polarizers.
The isolators are located approximately \unit{1}{m} away from the ultracold atoms which are susceptible to stray magnetic fields.
To overcome this problem the isolators are enclosed in mu-metal shielding.
Actuation of the power is performed after the optical isolators, such that the full fiber amplifier output continuously passes through them, minimizing thermal transients during the experimental sequence.

\section{Feedback System}

After the beam has been cleaned up and isolated, it is important to consider two aspects of the desired experiment:
\begin{enumerate}
\item The lattice power must be continuously tunable from around \unit{10}{\mu{}W} to around \unit{20}{W}, corresponding to lattice depths in the approximate range of $10^{-3}$ to $2 \times 10^3 E_R$, where $E_R$ is the geometric recoil energy corresponding to the lattice spacing~\cite{Blatt2015}.
\item The location of the minimum Gaussian beam waist must be accurately positioned to overlap the other lattices and dipole traps present in the experiment, and must remain aligned for the duration of the data collection process (hours to days).
\end{enumerate}
To satisfy the first of these requirements, a two-stage feedback system is implemented.
Two stages are used because the experiment typically operates in one of two regimes, which we term \emph{detection} and \emph{interaction}.
In the interaction phase of the experiment, the lattice is relatively shallow ($\approx \unit{100}{mW}$, corresponding to depths of $\approx 10 E_R$), which allows atoms to tunnel and interact with each other.
In this regime, precise and potentially fast control is required.
In the detection phase, the depth of the lattice is dramatically raised (to $\approx 2000 E_R$), which isolates atoms in their individual wells so that Raman sideband imaging can be performed~\cite{Parsons2015}.
In the detection regime, the control need not be fast, and the passive stability of the Mephisto laser ensures that noise is low at relevant frequency scales ($\approx\unit{2}{MHz}$ Fourier frequency).
Thus, two control loops are used, a fast loop based on an acousto-optic modulator that controls the laser at low powers, and a slow loop which uses a Berek compensator to control the laser at high powers.

\begin{figure}
  \begin{center}
    \includegraphics{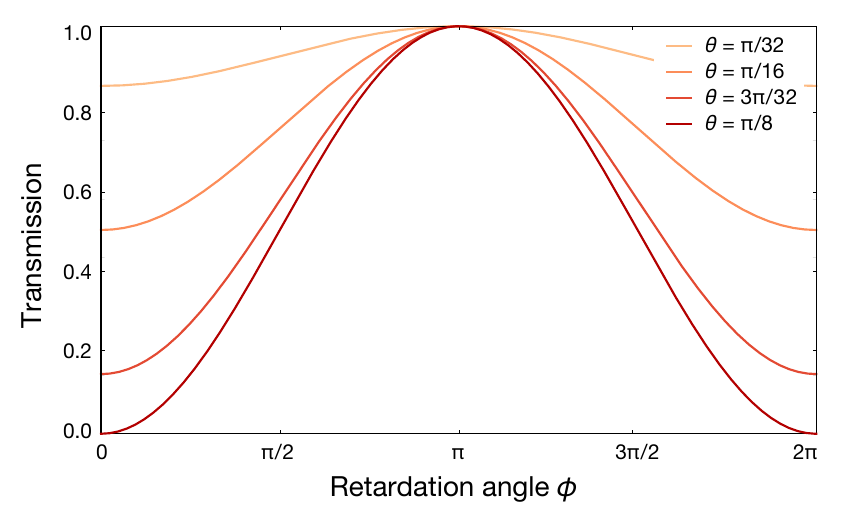}
    \caption{\textbf{Berek compensator} transmission as a function of the Berek compensator's retardation angle $\phi$ and waveplate rotation angles $\theta$.}\label{fig:berek}
  \end{center}
\end{figure}

After the isolation stage, we use the sequence of optical elements shown in Fig.~\ref{fig:optical_layout}, consisting of a polarizer, half-wave plate (HWP), Berek compensator mounted on a precision galvo (Thorlabs GVS002, or Camtech 8320K, depending on the beam path), half-wave plate, and polarizer.
The Berek compensator is a $z$-cut quartz plate anti-reflection-coated at \unit{1064}{nm}.
By tilting the plate about its vertical axis by a few degrees, the extraordinary axis is mixed into the propagation of the beam, leading to tilt-dependent birefringence that can be tuned from a zero-wave plate past a half-wave plate, represented by a retardation angle $\phi$.
Combined with the subsequent polarization optics, rotation of the Berek compensator can vary the transmitted power with a contrast of $\sim$1000:1.
Figure~\ref{fig:berek} shows the transmitted power as a function of retardation angle $\phi$ and fast-axis angle $\theta$ of the waveplates surrounding the compensator.

Full modulation contrast is achieved when $\theta=\pi/8$ with respect to the rotation axis of the Berek compensator.
This contrast can be reduced by varying $\theta$, to the limiting case where $\theta=0$, where the rotation of the Berek compensator does nothing, since the polarization of the light is parallel to the rotation axis of the quartz plate.

\begin{figure}
  \begin{center}
    \includegraphics[width=\columnwidth]{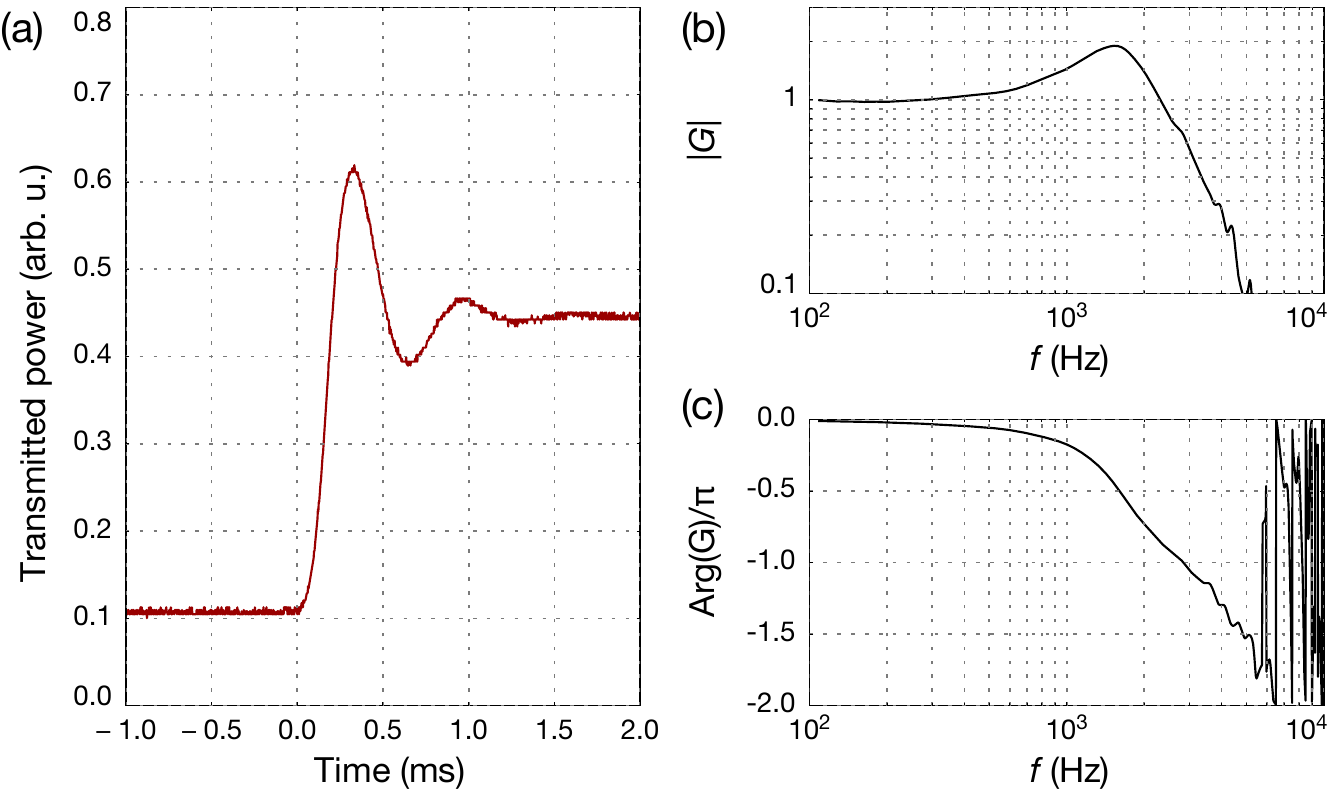}
    \caption{\textbf{Berek compensator based feedback loop}. This Berek compensator based feedback loop is used to vary the beam intensity at relatively high power. (a) Shows the small signal impulse response of the feedback loop (step size of $\approx 2$ W) and (b-c) show the corresponding Bode plots of the closed loop transfer function $G$.}\label{fig:berek_step_response}
  \end{center}
\end{figure}
A key point, as seen in Fig.~\ref{fig:berek}, is that if the waveplates are rotated together, the maximum transmitted power is fixed, while the minimum can be raised.
The waveplates are detuned such that the maximum power needed in \emph{interaction} mode is the minimum power transmitted by the Berek compensator system, strongly suppressing fluctuations due to angle positioning noise of the Berek compensator, since the slope of transmission with respect to the Berek angle vanishes at this point.
Naturally, rotation of a parallel plate in the beam path induces a shift of the beam, but this is acceptable because the Berek compensator is only active during the imaging phase of the experiment, when the position of the underlying harmonic trap is largely irrelevant.
The spatial phase of the optical lattice is set by the retro-reflector mirror, which does not move, and is not affected by the rotation of the Berek compensator.
This system functions as the actuator arm of a digital feedback loop, where the power of the beam is measured using a low noise photodiode. This feedback loop regulates the lattice between \unit{1}{W} of optical power and its maximum value (which depends on the amplifier, but is $\approx$\unit{20}{W}).
The step response and corresponding Bode plots~\cite{Bechhoefer2005} are shown in Fig.~\ref{fig:berek_step_response}.
A polarization rotation of $2 \pi$ is produced by approximately $3^\circ$ of axial rotation of the quartz plate.
Since we never need to rotate the plate outside this range, the position is constrained to this window in software.

Because two feedback loops are present in the system (the second one detailed in the next section), care must be taken so that they do not function simultaneously.
To accomplish this, the low-power loop is always manually railed to its high extreme when the high-power loop is active.
This is done by closing a mechanical shutter (Stanford Research Systems SR475) in front of the photodiode and setting the set-point to a high value, which quickly winds up the integrator to its maximum output.
Conversely, the digital high-power loop is disabled in software when the low-power loop is active.

\subsection{Low-power feedback}
The high-power feedback used in the \emph{imaging} phase of the experiment is exceedingly robust because the heating rate from scattered resonant imaging light far exceeds any technical heating from the loop~\cite{Blatt2015}.
Even if the loop was noisy, the heating would be entirely compensated by the simultaneous Raman cooling~\cite{Parsons2015}.

In contrast, the low-power loop that controls the lattice laser during the \emph{interaction} phase of the experiment is critically important:
\begin{enumerate}
  \item Its effect on the RIN can limit the system temperature and dominate heating rates, limiting the amount of time available for atom interaction~\cite{Blatt2015}.
  \item It needs to avoid varying the beam pointing because the trap position matters during this phase of the experiment.
  \item It must have closed-loop bandwidths $\geq\unit{1}{kHz}$, and open-loop modulation bandwidths $\geq \unit{100}{kHz}$.
\end{enumerate}
To this end, we use an nonlinear feedback loop based on a TeO$_2$ AOM (Gooch \& Housego AOMO 3080-198), where the power of the RF supplied to the AOM is used to actuate the fast, low-power feedback loop.
The AOM was chosen for its low thermal lensing and easily-accessible RF powers.
It is mounted on a monolithic, water-cooled flexure mount that allows optimization of the AOM efficiency, while remaining stable over years of operation.

\begin{figure}
  \begin{center}
    \includegraphics{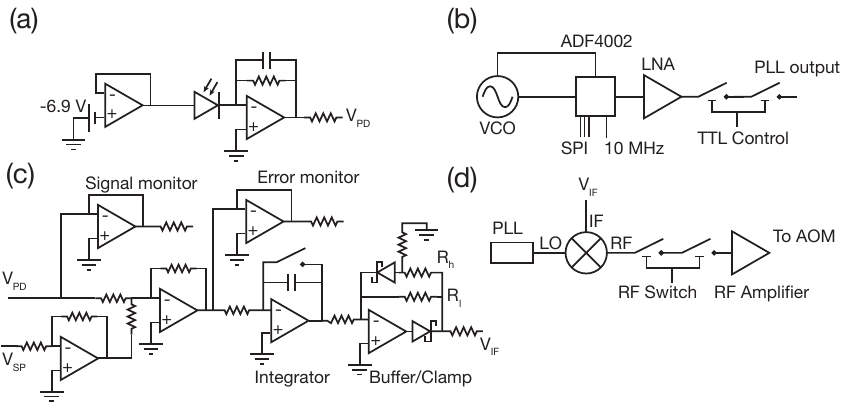}
    \caption{\textbf{Feedback circuits.} (a) The reverse-biased photodiode. (b) Externally referenced phase-locked loop. (c) Nonlinear loop filter. (d) RF power modulation using a mixer.}\label{fig:circuits}
  \end{center}
\end{figure}
The feedback circuit is shown in Fig.~\ref{fig:circuits}(c). As seen in Fig.~\ref{fig:optical_layout}, a pair of low-noise photodiodes [Fig.~\ref{fig:circuits}(a)] measures the optical power of the beam sampled with a weakly reflective optic (Thorlabs BSF10-C).
The beam sampler is wedged in such a way that the two reflections diverge from one another, allowing us to illuminate both photodiodes.
One of the sampled beams is attenuated by an additional factor of 20 by reflection off of an uncoated fused silica glass plate, such that the dynamic range of its photodiode covers \unit{0-20}{W} of the beam directed at the atoms, and controls the Berek-compensator-based, high-power feedback loop.
The other beam is not attenuated, save by the beam sampler, and its dynamic range covers \unit{0-1}{W}.
%To prevent damage to the low-power photodiode when the system is at its highest powers, a shutter (Stanford Research Systems SR475) blocks the light when the high-power loop is in use.
The low-power photodiode passes its signal $V_{\mathrm{pd}}$ to a loop filter circuit [Fig.~\ref{fig:circuits}(c)], which compares it to the control system's set-point, generated by a digital-to-analog converter.
The loop filter's output controls the IF port of a microwave mixer [Minicircuits ZFM-2-S+, see Fig.~\ref{fig:circuits}(d)], to modulate the RF power supplied by a low-noise phase-locked loop [PLL, see Fig.~\ref{fig:circuits}(b)].
The resulting RF signal is then amplified and sent to the AOM.
Let us now consider every part of this signal chain individually.

\subsubsection{Phase-locked loop (PLL)}
The local oscillator (LO) RF source is a custom-built printed circuit board (PCB), schematically shown in Fig.~\ref{fig:circuits}(b), that implements a phase locked loop (PLL) based on the ADF4002 IC and a low-phase-noise voltage-controlled oscillator (VCO, Crystek CVCO55CL-0060-0110), preamplified with a low-noise amplifier (Minicircuits PSA4-5043+).
The PLL is locked to the \unit{10}{MHz} frequency reference distributed throughout the lab to ensure absolute frequency stability.
In addition to PLL functionality, the board also contains two digitally-controlled, high-isolation RF switches, giving a total RF isolation exceeding \unit{90}{dB}.
This device allows us to drive the AOM with a single, clean tone ($<$\unit{-90}{dBc/Hz} at \unit{200}{kHz} away from the carrier).
%The spectrum and relative intensity noise of the PLL can be found in Fig.~\ref{fig:pll}.
Thus, the spectrum of the PLL exhibits phase noise comparable to that of a typical VCO but without the slow frequency drift characteristic to these devices.
Elimination of the carrier drift is worth the added phase noise because any drift of the frequency affects the pointing of the laser beam passing through the AOM, which would misalign the optical lattice.

\subsubsection{Photodiode}

\begin{figure}
  \begin{center}
    \includegraphics[width=\columnwidth]{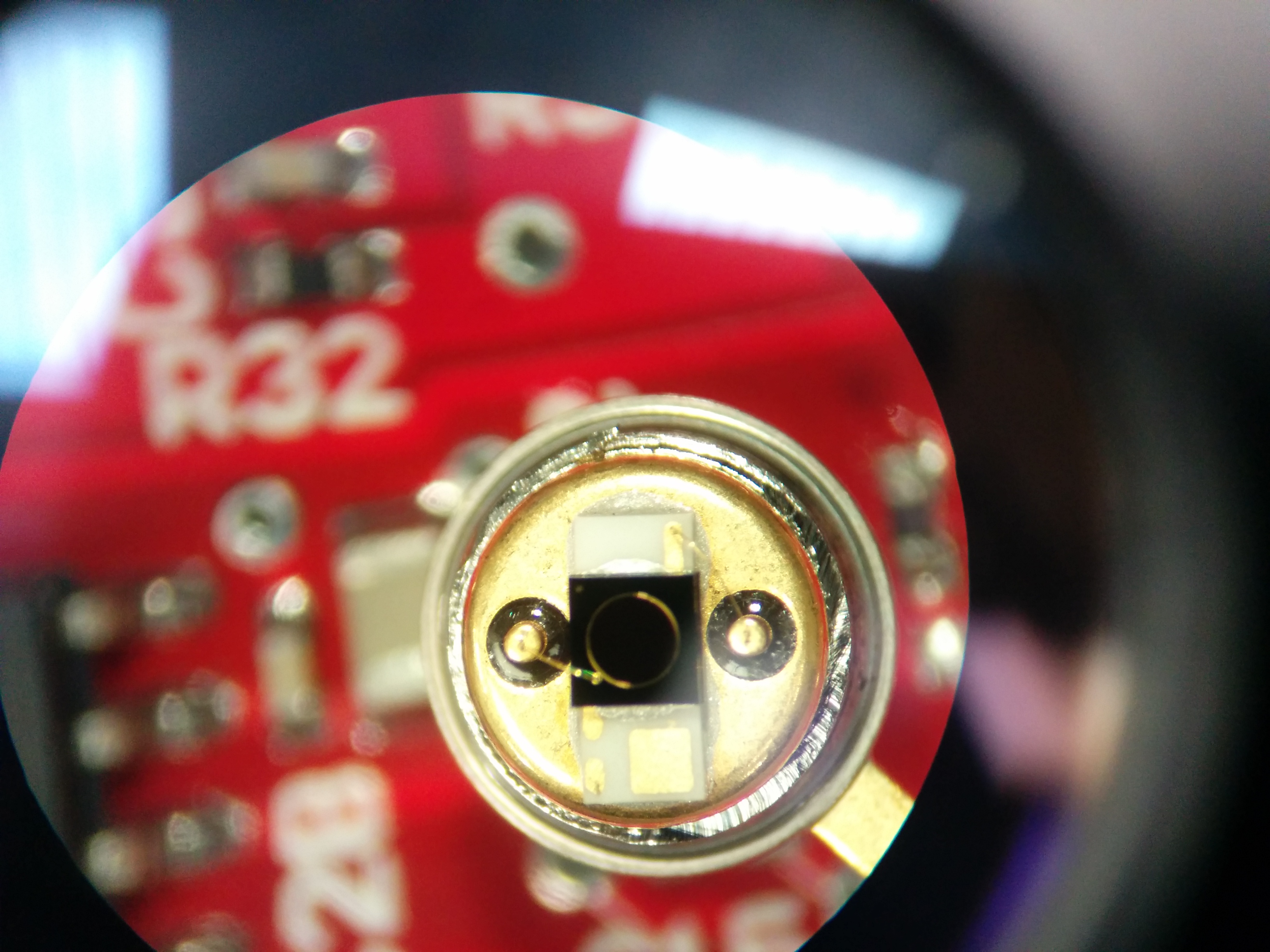}
    \caption{\textbf{Photodiode photo} showing that the ``can'' of the photodiode has been cut, to suppress etalon-like behavior between the cover glass and the sensor.}\label{fig:cut_pd}
  \end{center}
\end{figure}

Fast, low-noise photodiodes are required both to provide the monitoring arm of the feedback loop and to characterize the RIN of the system.
An ideal device would have a bandwidth up to approximately \unit{3}{MHz} (since trap frequencies in the system are \unit{1.5}{MHz}), with a noise floor below approximately \unit{-150}{dBc/Hz} over that range~\cite{Blatt2015}.
To approach these specifications, we use a custom-designed PCB that implements a simple transimpedance amplifier~\cite{Graeme1995} in a convenient physical package, providing bandwidths in the \unit{0.1-10}{MHz} range, depending on the transimpedance gain set by a resistor.
The photodiode can be safely operated up to incident powers of approximately \unit{10}{mW}, which sets a \unit{-164}{dBc/Hz} shot noise limit on the RIN that can be detected.
In practice, we typically operate the photodiodes with $\approx$\unit{1-5}{mW} of incident power.
A photodiode with a bandwidth (BW) of \unit{6}{MHz} is used to characterize the system, while photodiodes with a bandwidth of \unit{330}{kHz} (and correspondingly higher gain) are used to perform the feedback.

The chosen photodiode (Hamamatsu G8370-01) is an InGaAs based, small-area, low-noise design with a high efficiency and low thermal sensitivity at \unit{1064}{nm}.
Prior to use, the protective window is removed from the photodiode to prevent interference effects, as seen in Fig.~\ref{fig:cut_pd}.
Further, to reduce the capacitance of the system, a reverse bias of \unit{6.9}{V} is applied to the photodiode by a temperature-stabilized voltage reference (Linear Technology LM399H).
The transimpedance circuit is shown in Fig.~\ref{fig:circuits}(a).
It converts the photo current supplied by the photodiode to a voltage signal using a low-noise and high-bandwidth op-amp (Texas Instruments OPA843).
To reduce electrical pickup and prevent ground loops, the electronic assembly is housed in a metal enclosure and mounted to the optical table using electrically insulating mica posts.

Light is focused onto the photodiode using an $f=\unit{25.4}{mm}$ lens such that the beam at the photodiode is significantly smaller than the photodiode diameter, to prevent pointing fluctuations from registering as amplitude fluctuations.
An iris and interference filter (Semrock FF01-1020/LP25) are used to eliminate stray sources of light.

\subsubsection{Digital-to-analog converter}
\begin{figure}
  \begin{center}
    \includegraphics{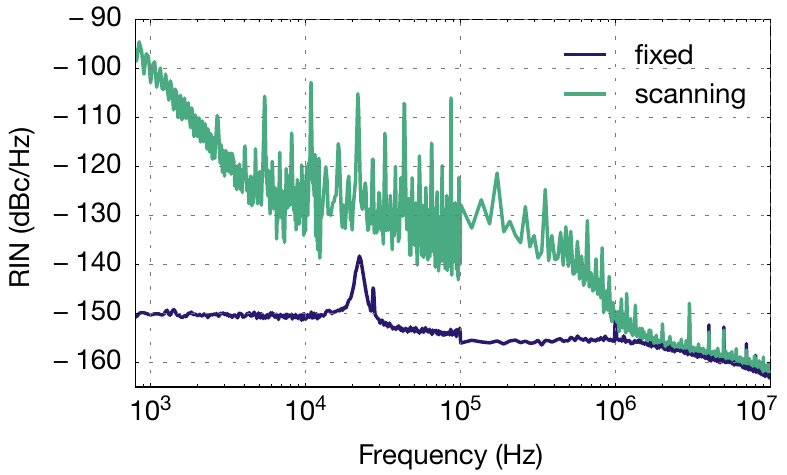}
    \caption{\textbf{Relative intensity noise of the digital to analog converter} at a fixed output voltage and while changing values. To measure the transient RIN, a triangle wave with \unit{20}{ms} period, a DC offset, and a modulation depth of 15\% of the DC offset was produced by the device. The triangle wave accounts for the series of spurs at low frequencies that fall with increasing frequencies. At frequencies above approximately \unit{4}{kHz}, the spectrum is dominated by noise due to transient ``glitches'' that emerge when the device changes its output state.}\label{fig:da_rin}
  \end{center}
\end{figure}

The digital-to-analog converter (DAC, Texas Instruments DAC9881) connects the experimental digital control system to the analog control loop for the lattice.
By definition, an ideal control loop would translate any noise on the analog signal from the DAC to the laser light, meaning that a poorly designed DAC would inherently limit the lattice laser performance.
The noise on the DAC, when it is programmed to output a constant voltage is shown in Fig.~\ref{fig:da_rin}, and is found to be roughly constant at a level of \unit{-150}{dBc/Hz} --- more than acceptable.
When the state of the DAC is changed, such as during a ramp, this noise level may be higher, due to the so-called ``glitch'' energy, as seen in the figure.
This transient noise is mitigated by two factors: (1) it is only present when the lattice power is changing, which does not happen for most of the sequence, and a lattice at fixed power will have negligible RIN contributions. Further, (2) the loop filter strongly reduces the noise at higher Fourier frequencies.

\subsubsection{Loop filter}

The connection between the actuation and monitoring mechanisms of the feedback system is the loop filter, which compares the \emph{set-point} signal, $V_\mathrm{SP}(t)$, supplied by the experimental control system and the photodiode signals $V_\mathrm{PD}(t)$, to adjust the actuation accordingly.
We use a simple integrating design in three stages, as seen in Fig.~\ref{fig:circuits}(c), based on a low-noise op-amp (Texas Instruments OPA211).
The first stage computes and amplifies an error signal, $e(t)=V_\mathrm{SP}(t)-V_\mathrm{PD}(t)$, which is integrated by the second stage.
The integral gain is set using a low-drift trim-pot.
This is the case because within relevant frequency ranges, the AOM can be coarsely approximated by a pure time delay process, given by the acoustic wave transit time of $\approx$\unit{1}{\mu{}s} across the active region of the AOM crystal.
In this scenario, the proportional term must be small to avoid instability at the $\pi$ phase shift point, and a pure integral controller is optimal~\cite{Skogestad2003}.
For debugging purposes, the loop filter features buffered photodiode monitor and error signal monitor outputs.

The third stage of the loop filter is used to accommodate the fact that the actuation arm is nonlinear: the optical power transmitted by the AOM is proportional to the RF power incident on it $P_{\mathrm{rf}}\propto V_{IF}^2$, where $V_{IF}$ is the actuation signal supplied to the IF port of the mixer.
In effect, if the rest of the feedback system was linear, this would correspond to a quadratically increasing gain until the AOM is saturated, which leads to slow behavior at low signal levels and ringing at higher signal levels.
Even worse, since the mixer is insensitive to the sign of the voltage applied, the gain is negative for negative voltages, resulting in an unstable system.

\begin{figure}
  \begin{center}
    \includegraphics{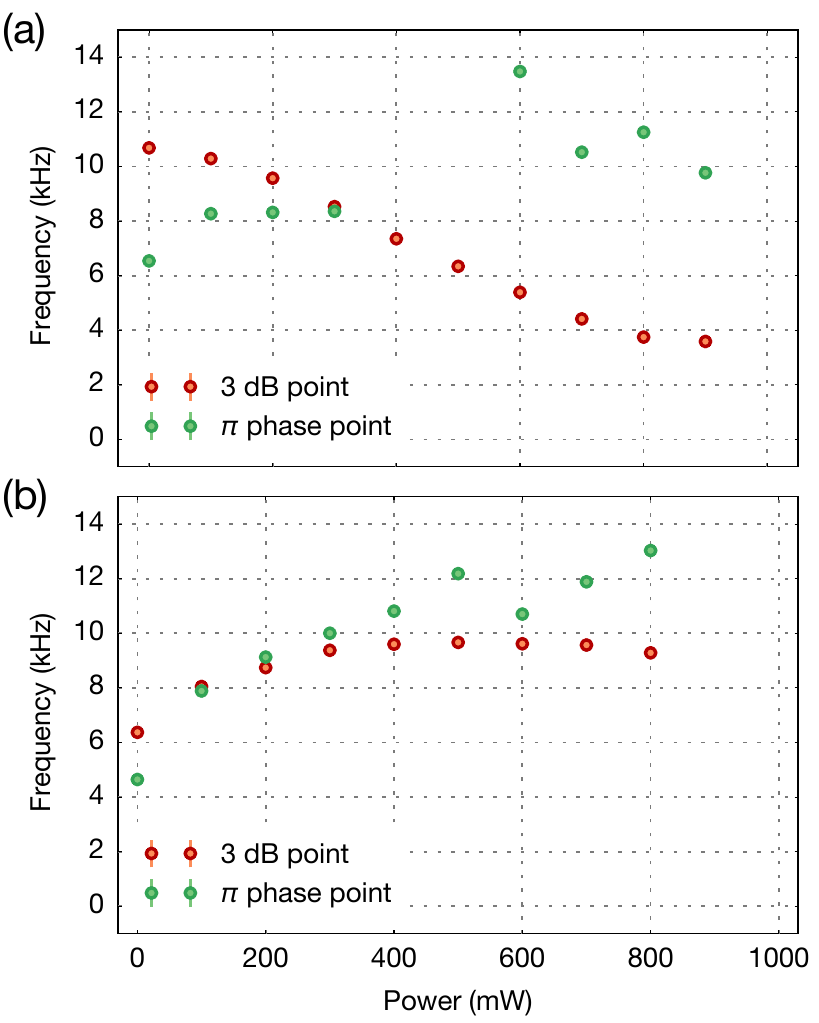}
    \caption{\textbf{Loop bandwidth dependence on the set-point}. (a) shows the \unit{3}{dB} and $\pi$ phase shift points for the $X$ lattice. (b) shows the same for the $Y$ lattice.}\label{fig:bandwidth}
  \end{center}
\end{figure}

To achieve the desired functionality, the final op-amp stage linearizes the response by increasing the gain at low signal levels and suppressing it at high signal levels by including a fast Schottky diode (ON Semiconductor BAT54S) in the feedback arm~\cite{Horowitz2015}.
This can be seen by considering two extreme cases: far below the diode drop, the diode is an open circuit and the gain is set by $R_l$; above the diode drop it is a closed circuit where the gain is set by the resistor $R_h$.
Naturally, the voltage drop across the diode is chosen such that it is approximately half the maximum desired signal level of the photodiode (\unit{3.3}{V}).
Although a large improvement, the bandwidth still varies over the dynamic range of the loop, but only by approximately a factor of two, as seen in Fig.~\ref{fig:bandwidth}.
Were it not for the linearization, the gain of the mixer would change by a factor of 10 over a dynamic range spanning two orders of magnitude.
The loop transfer function is sensitive to the specific tuning parameters of the nonlinear controller resulting in slightly different transfer functions for the two axes, although both exhibit approximately constant behavior over the desired range of setpoints.
Another function of the last stage is to clamp the voltage to be strictly positive by including a diode clamp on the output to prevent unstable positive feedback behavior~\cite{Horowitz2015}.
Finally, since the mixer is a current driven device, the last stage of the feedback loop uses the low-noise OPA627 (Texas Instruments) op-amp which is able to supply tens of mA of current into an input impedance of \unit{50}{\Omega}.

\begin{figure*}
  \begin{center}
    \includegraphics{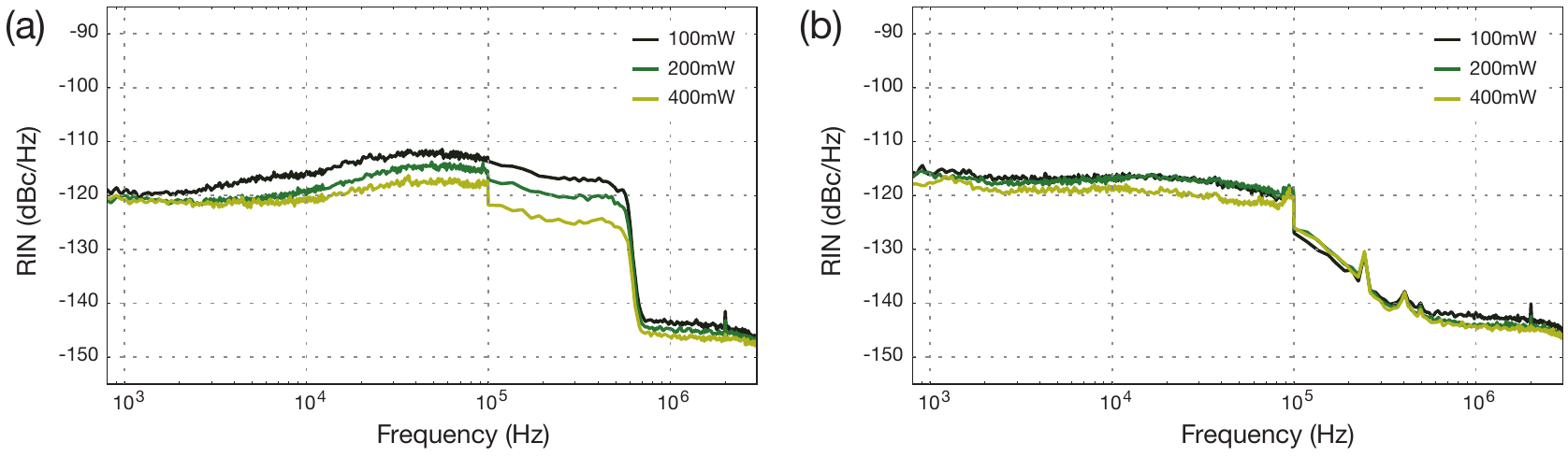}
    \caption{\textbf{Laser noise of the lattice system with and without additional low pass filtering.} (a) Closed-loop low-power server using a different fiber amplifier than the one in Fig.~\ref{fig:noises}(d). (b) Noise on the same system as in panel (a), but with additional low-pass filters.}\label{fig:low_pass_noises}
  \end{center}
\end{figure*}

A common undesired behavior of analog feedback systems with integral gain is \emph{integrator wind-up}~\cite{Bechhoefer2005}, which can occur when the feedback loop is manually broken, meaning that the actuator can no longer adjust the state of the controlled parameter.
In this regime, the integrator accumulates a large positive or negative voltage, such that when the loop is mended, the system briefly rails before control is restored.
In our case, \emph{integrator wind-up} occurs when the high-isolation RF switch is open during the state preparation procedure, and closed when the optical lattice is desired.
In most cases, the optical lattice must be turned on adiabatically, to avoid heating the trapped atoms due to fast transients resulting from integrator windup.
To solve this issue, a bypass switch (Maxim MAX4503) is placed across the capacitor of the second stage of the feedback circuit, which siphons off the charge when the lattice is inactive.
Even when the switch is closed, its resistance is nonzero (due to a protection resistor that limits the current that can pass through the switch), the conditions leading to integral windup in an unprotected system still lead to charge accumulation on the integral capacitor, albeit at a much reduced level.
Thus, in addition to the switch, when the system is inactive, the set-point is set to the dark signal from the photodiode ($\approx$\unit{200}{\mu{}V}, calibrated specifically to each photodiode), making the error signal very small.

To avoid a so-called ``servo bump'' in the noise spectrum, and because there is little need for fast, closed-loop control, the loop filter is tuned very conservatively, to a bandwidth of approximately \unit{10}{kHz} over most of the range.
However, a small amount of tuning can increase the bandwidth up to $\approx$\unit{150}{kHz}, which is limited by the propagation delay of the acoustic wave in the AOM.
Although fast, closed-loop control is not available in the current configuration, the last stage of the loop filter features an adder connected to an optional ``feed-forward'' port.
This port can be used to modulate the RF power at frequencies above the bandwidth.

Although it is an essential part of the system, the loop filter is also one of its major limitations. Figure~\ref{fig:noises} shows the noise performance of the output of the fiber amplifiers in open-loop compared to the noise in closed-loop.
Noise near DC is lowered (as expected from a closed-loop system), noise outside the loop bandwidth is increased in the range of 10 to \unit{450}{kHz} as seen in subfigures (c) and (d).
This can be explained by the noise gain of the op-amp circuits in combination with the speed limitations of the nonlinear output stage~\cite{Horowitz2015}.
The sharp drop off at the \unit{450}{kHz} point is due to a low-pass filter placed on the output of the loop filter, far above the maximum bandwidth set by the AOM.
When the broadband noise below \unit{450}{kHz} was determined to be a problem, we decided that a bandwidth of higher than \unit{10}{kHz} was not required in the foreseeable future.
As a result, we added a first-order low pass with a \unit{3}{dB} frequency of \unit{10}{kHz} leading to the improvement shown in Fig.~\ref{fig:low_pass_noises}. The fiber amplifier used to take this data was different from the one used for Fig.~\ref{fig:noises}, although their performance is comparable.

\subsection{Beam shaping}
After the beam has passed through the feedback optics, it must be applied to the atoms. To that end, we use a $2f$ telescope of short focal length triplet collimator lenses to expand the beam, and then focus it with a long focal length singlet lens.
We focus the lattice to a \unit{70-90}{\mu{}m} waist at the position of the atoms.
To verify that the waist is positioned correctly, we use a temporary mirror to redirect the beam away from the vacuum chamber (a glass cell) and onto a camera positioned in the plane corresponding to the position of the atoms.

To prevent unwanted interference, the lattice beams enter the glass cell at an angle.
Unfortunately, due to the angle of incidence, the beam contains both $S$ and $P$ polarizations, which are reflected differently at the glass cell wall, thus altering the polarization state of the beam.
Even worse, this happens on both the initial and retro-reflected pass through the glass cell.
For this reason, we use a combination of quarter waveplate (QWP) and half waveplate (HWP) to optimize the interference contrast of each lattice axis.
This is done by optimizing the on-site trap frequency of lattice, measured using lattice-modulation spectroscopy~\cite{Blatt2015}.

\section{Beam monitoring and alignment}
Improved spatial resolution and control have given experimentalists the ability to use relatively few atoms~\cite{Preiss2015, Choi2016, Mazurenko2017} that are simultaneously addressed by multiple beams and held by multiple traps.
Such experiments place stringent bounds on the required beam pointing stability.

\begin{figure*}
  \begin{center}
    \includegraphics{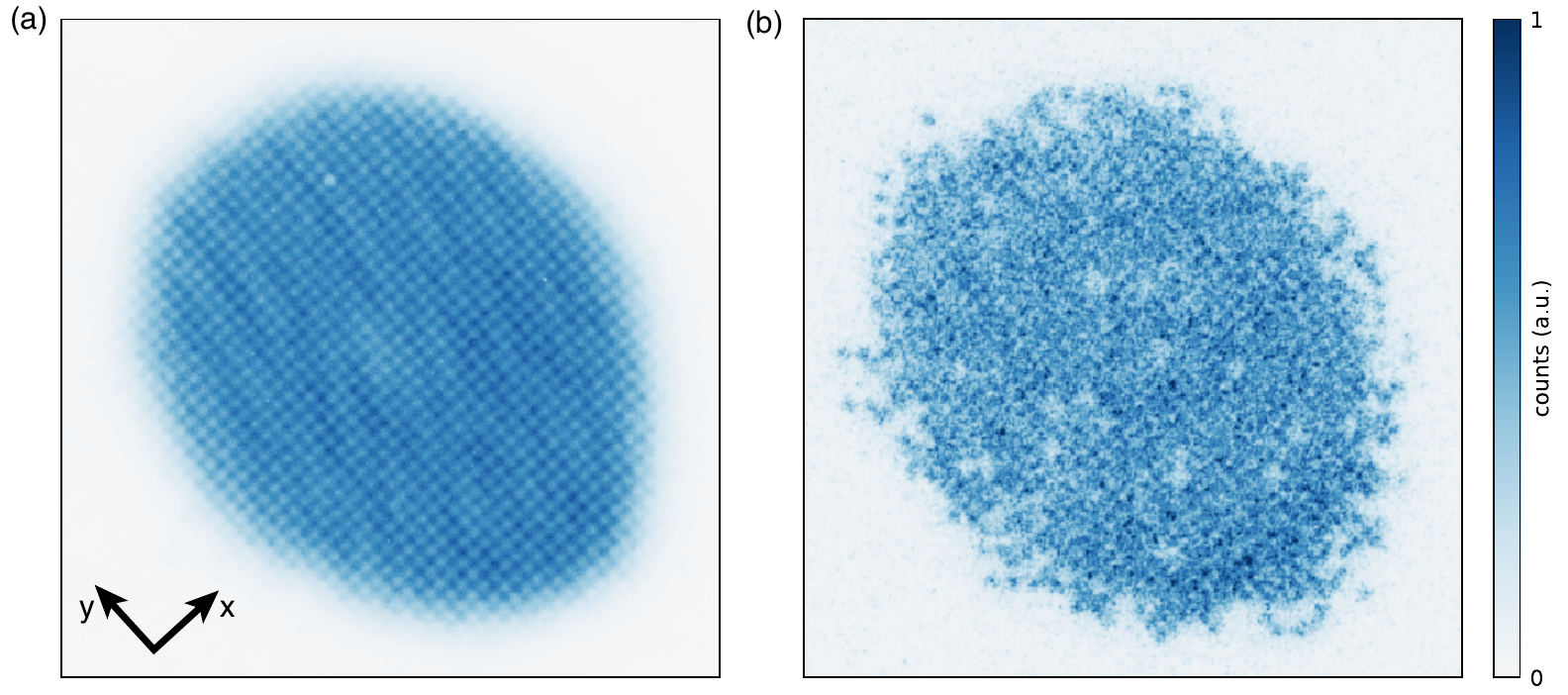}
    \caption{\textbf{Averaged images of Mott insulators}. (a) Several Mott insulators were created over the course of approximately one hour, and the average image is shown~\cite{Greif2016}. If the lattice pattern would have fluctuated by more than a small fraction of a lattice site, the modulation contrast would have been washed out. The absence of this effect means that the phase of the optical lattice is stable. (b) shows an individual image of a Mott insulator.}\label{fig:averaged_mott}
  \end{center}
\end{figure*}

Several applications require the positions of lattice sites to not change by more than a small fraction ($\approx 10\%$ of the lattice constant) between experimental shots.
In our setup this is ensured by careful design of the optical assembly, where the phase of the retro-reflected lattice is set by the retro-reflecting mirror, which is placed approximately \unit{5}{cm} from the atoms~\cite{Huber2014}.
This phase determines the location of the sites of the optical lattice.
We have characterized the phase stability of the optical lattice by recording a long sequence of densely populated images with our imaging camera. We found that the phase varies by much less than one lattice site over the course of hours, as seen in Fig.~\ref{fig:averaged_mott}.

The position of the atomic cloud, however, is not determined by the phase difference of the retro-reflecting lattice beams, but by the underlying harmonic confinement generated by their transverse Gaussian profiles.
The stability of this confinement is governed by the pointing stability of the optical lattice laser beams.
In experiments where the entire cloud is addressed by beams auxiliary to the lattice, which can be used for example for enhanced entropy redistribution~\cite{Mazurenko2017}, reducing drifts and fluctuations of the lattice position is important.

The required positional stability of the underlying harmonic trap is achieved through a combination of passive stability and active monitoring.
Passive stability is ensured by using stable optomechanics (largely from the Thorlabs Polaris line), mounted on short optical posts (\unit{2}{inch} beam height relative to the floated optical table).
To limit sensitivity to air currents and further improve stability, the beam is enclosed with \unit{1}{inch} aluminum tubes throughout its path and the entire setup is covered by metal plates.
As a final measure to isolate the experimental setup, the optical table is contained within an aluminum enclosure. As detailed in prior sections, care has been taken to minimize thermal lensing and drift in optics encountered by the lattice beam.

Generally, changes in environmental conditions (such as temperature and relative humidity), load on the water-cooling systems and other external conditions can lead to a slow drift in the laser pointing over long timescales.
The situation is further complicated by the fact that opening the enclosure of the experiment changes the thermal state of the system.
For these reasons, each lattice axis is fitted with a remotely actuated mirror, controlled by the experimental control software over the lab network, allowing for non-invasive, digital alignment of the system.
The remote actuators are ``Picomotors'' (Newport 8302), fitted into customized \unit{2}{inch} mirror mounts.
They work in a ``fire-and-forget'' manner, remaining passively stable when not in active use.

Variation in external conditions can contaminate large datasets if the prepared state sensitively depends on laser pointing.
To guard against such contamination, we have implemented an independent monitoring system to verify appropriate pointing of both lattice laser beams.
This monitoring system allows us to eliminate (via post-selection) experimental realizations during which the system was in an undesired state.
A beam-sampler (Thorlabs BSF10-C) reflects a small portion of the beam shortly before it reaches the atoms and images it onto a monitoring camera located in the plane corresponding to the atomic system, imaging a Gaussian beam corresponding to the optical lattice.
Naturally, the position of the atoms is dictated by the light field of both the incoming and retroreflected beams, while the monitoring camera only observes the position of the incoming beam.

In our experiments, the position of the Gaussian beam envelope of the optical lattice is impossible to determine from the site-resolved image of the atoms, because the atoms may not occupy the central part of the envelope depending on the specifics of the lattice loading procedure.
In addition, the atomic distributions are governed by the sum of the incident lattices and thus cannot be used to reliably infer the degree of overlap between the lattice beams.
Nevertheless, it is possible to adiabatically load a strongly anisotropic optical lattice by lowering the power of one of the lattice beams, resulting in an atomic distribution that closely matches the trap produced by the high-power beam~\cite{Parsons2015}.
This procedure yields the correct position of the Gaussian beam envelope in the atom plane, which we can use to analyze each beam's passive stability and its relationship to the beam-monitoring camera.

\begin{figure}
  \begin{center}
    \includegraphics[width=\columnwidth]{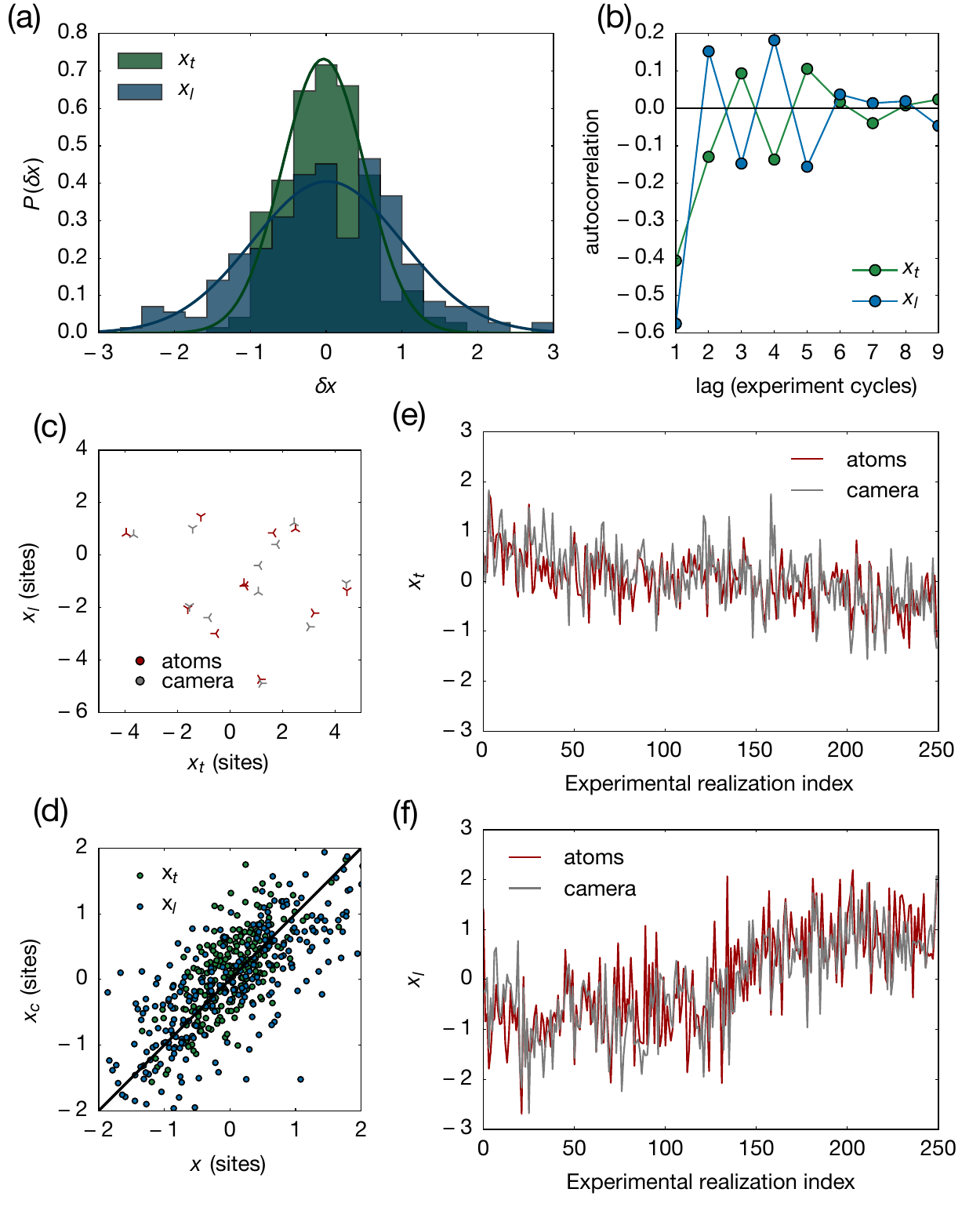}
    \caption{\textbf{Short term positional stability.} (a) A histogram of shot-to-shot step sizes along the transverse and longitudinal directions, with a Gaussian fit. (b) Correlations between the step size of subsequent shots. (c) Calibration data for the ``affine fit'' of the conversion from camera pixels to lattice dimensions. (d) Predicted versus measured location along the transverse and longitudinal directions. (e,f) Measured  and predicted position of the atomic cloud, based on refitting the offset every $20$ experimental realizations.}\label{fig:stability}
  \end{center}
\end{figure}

We first characterize the passive stability of the optical lattice along one dimension by monitoring its position over the course of 250 experimental realizations.
We decompose the lattice position into the transverse $x_t$ and the longitudinal $x_l$ components with respect to the optical axis.
Figure~\ref{fig:stability}(a) shows a histogram of displacements of atoms between subsequent experimental realizations, showing that the RMS displacement is $0.54(3)$ [$0.98(4)$] sites along the $x_t$ [$x_l$] direction.
The variation of the displacements is a geometric effect consistent with the mechanical assembly of the experiment~\cite{Huber2014}.
The distributions are centered within 1-$\sigma$ of zero along both directions: $-0.03(4)$ and $0.01(5)$ sites for transverse and longitudinal axes, respectively.
If the trap had followed a random walk with these step sizes, the optical lattice would have to be realigned frequently, since after only $100$ experimental shots, the trap would have moved by more than $10$ lattice sites.
However, the situation is more fortunate, since subsequent steps are correlated, as shown in Fig.~\ref{fig:stability}(b).
Once this correlation is taken into account, we estimate an RMS drift of 8(1)~sites over 1000~realizations in the transverse direction and 12(2)~sites over 1000~realizations in the axial direction.
This drift corresponds to an RMS displacement of $\unit{14(3)\times 10^{-3}}{sites / realization}$, compared to a beam waist of $\approx$\unit{80}{\mu{}m} ($\approx$140 sites).

In general, we cannot extract the lattice position from the atomic distribution directly, but rely on the monitor camera picture, which is related to the lattice position in the atom plane by an affine transformation, consisting of a rotation, a scaling, and an offset.
To determine the parameters of this affine transform, we use the remote actuators to purposely misalign the lattice beams and correlate the monitor camera image with the atom image obtained by fluorescence imaging~\cite{Parsons2015}.
The corresponding least-squares fit is shown in Fig.~\ref{fig:stability}(c).
The quality of this fit remains acceptable for $\approx$30 repetitions of the experiment, but it degrades for longer times.
This effect is most likely due to thermal drift of the beam-monitoring camera, since a displacement of a one pixel (\unit{3}{\mu{}m}) on the monitoring camera is equivalent to a displacement of one lattice site in the atom plane.
The most likely parameter of the affine transform susceptible to thermal drift over a few experimental cycles is the offset.

\begin{figure}
\begin{center}
    \includegraphics[width=\columnwidth]{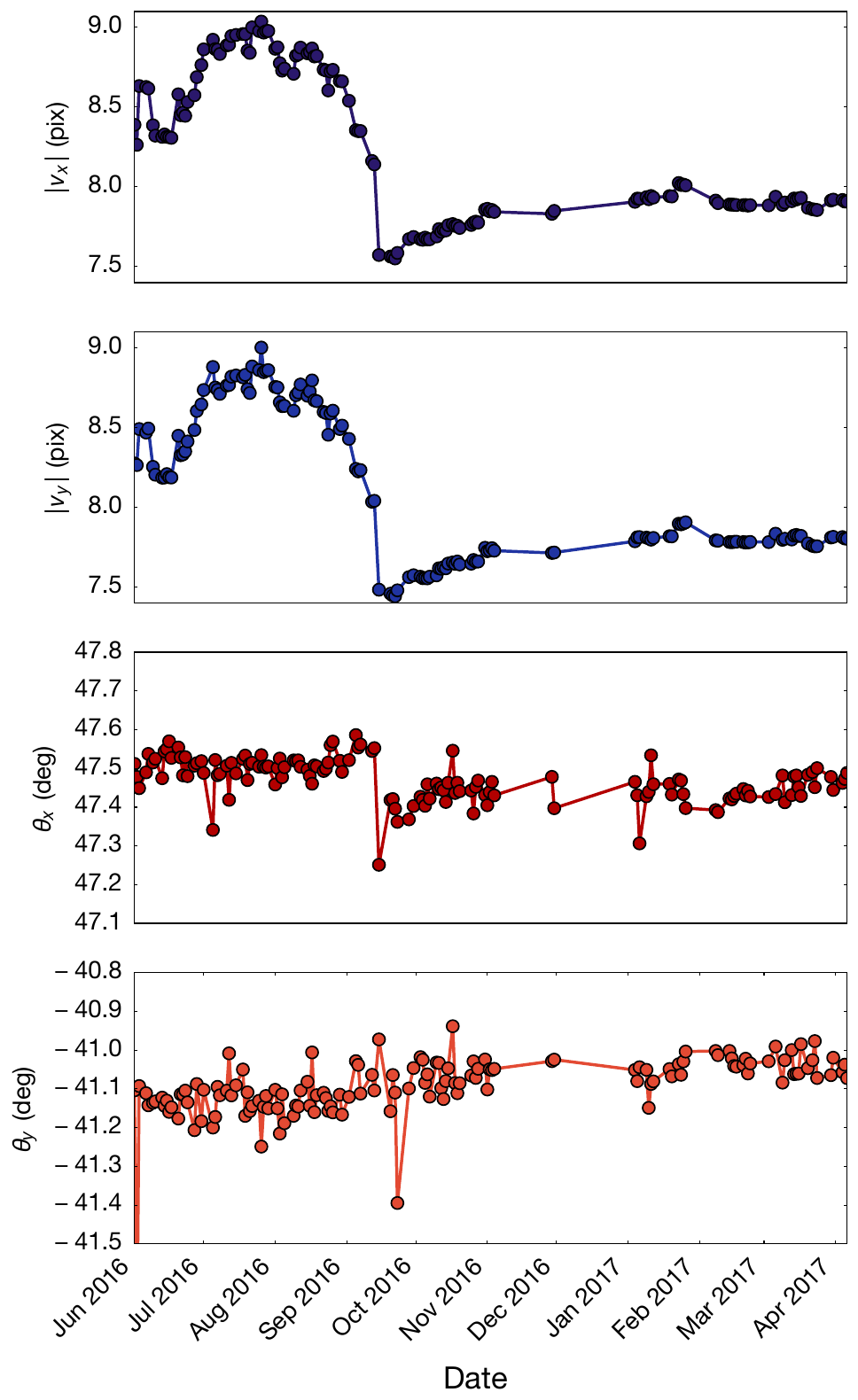}
    \caption{\textbf{Long term stability over approximately ten months.} The lattice spacings $|\mathrm{v}_x|$ and $|\mathrm{v}_y|$ and the lattice angles $\theta_x$ and $\theta_y$ along the two lattice axes $x$ and $y$ (cf. Fig.~\ref{fig:averaged_mott}) are shown over a period of ten months.}\label{fig:long_term_stability}
  \end{center}
\end{figure}

For this reason, we developed a procedure to periodically update the affine transform relating the beam-monitoring camera to the atom plane.
To test this procedure, we analyze the data set of monitoring-camera measurements and atomic positions (shown in Fig.~\ref{fig:stability}) in the following way: we split the data set into sequential blocks of 20 shots, each followed by a block of 8 shots.
We try to predict the larger blocks, by analyzing the smaller blocks to derive the updates to the affine transform.
The results are shown in Fig.~\ref{fig:stability}(d-f) and display a large degree of tracking between the predicted and measured atom positions.
The average taxicab distance between the predicted position and the measured position is 0.7~lattice sites.
This result shows that recalibrating the affine transformation every 20 shots bounds the tracking error to an acceptable level. In practical applications, larger blocks of shots containing scientific measurements are interleaved with small recalibration blocks.

In conclusion: the beam-monitoring camera is sufficiently sensitive to detect lattice misalignments on the order of a few lattice sites, making it useful as a noninvasive check of the lattice position.

\section{Long-term stability of the lattice structure}
A benefit of site-resolved imaging is that every measurement of the atomic distribution is also a direct measurement of the underlying lattice structure (except for the short-term and small-scale displacement discussed in the last section).
As discussed in the supplementary material of~\cite{Parsons2016, Greif2016}, the Fourier transform of an atomic distribution contains distinct peaks corresponding to the lattice structure, and these peaks can be used to extract the relative angle and the spacing of the lattices.
For this reason, we recorded a long-term measurement spanning almost one year to learn how these parameters change over time, see Fig.~\ref{fig:long_term_stability}.
Several interesting features can be seen from the observed behavior. In the first few months the apparent lattice spacing (as imaged through the microscope) changed by around 10\%. Experimentally, this drift was accompanied by the necessity of refocusing the imaging system. The sudden jump in September 2016 was caused by a simplification of the imaging system. Afterwards, the lattice spacing seemed more stable (fractional changes of $\leq 0.005$ and limited by the imaging system magnification drift). Unlike the lattice spacing, the angle is observed to be mostly unaffected and remains very stable (for example, an RMS of 0.06$^\circ$ in Spring 2017).

\section{Conclusion}
We have presented the design and specifications of an optical lattice system that has been successfully used for quantum gas microscopy of low-temperature phases of the Hubbard model.
In this design, key emphasis has been placed on the stability of the pointing and the intensity of the optical lattice, of particular importance to observe the underlying physics.
To further improve the system's performance, the system can be upgraded in a variety of ways. For example, the Nufern fiber amplifiers can be replaced with a lower noise model [such as the MOPA system shown in Fig.~\ref{fig:noises}(a)]. The nonlinear analog feedback loop can also be replaced by a lower-noise digital version.
Active stabilization of the laser pointing can improve the pointing stability further.
The high degree of lattice stability demonstrated here has already enabled first investigations of entropy redistribution schemes involving multiple traps that made it possible to observe antiferromagnetic ordering in two-dimensional Fermi gases~\cite{Mazurenko2017} and to engineer low-entropy quantum states~\cite{Chiu2018}.

% \bibliography{latticeoptics}


\end{document}